\newcounter{myctr}
\def\myitem{\refstepcounter{myctr}\bibfont\noindent\ifnum\themyctr>9\else\phantom{0}\fi\hangindent17pt\themyctr.\enskip}

\documentclass{ws-ijqi}
\usepackage[super,sort,compress]{cite}
\usepackage{xcolor}
\usepackage{braket}
\usepackage[verbose,hypertexnames=false]{hyperref}
\hypersetup{colorlinks=false,allbordercolors=blue,pdfborderstyle={/S/U/W 1}}

\newcommand{\cop}[0]{\gamma \,}
\newcommand{\te}[0]{\theta\,}

\usepackage{ulem}
\usepackage{color}
\definecolor{lavender}{rgb}{0.71, 0.49, 0.86}

\begin{document}

\markboth{Luciano Manara, Andrea Smirne and Bassano Vacchini}{Non-Markovianity in the Adapted Caldeira-Leggett model}

\catchline{}{}{}{}{}

\title{Non-Markovianity in the Adapted Caldeira-Leggett model}

\author{Luciano Manara }

\address{Istituto Nazionale di Fisica Nucleare, Sezione di Milano, Via  Celoria 16, I-20133 Milan, Italy\\
luciano.manara@mi.infn.it}

\author{Andrea Smirne}

\address{Università degli Studi di Milano, Dipartimento di Fisica, Via  Celoria 16, I-20133 Milan, Italy\\
Istituto Nazionale di Fisica Nucleare, Sezione di Milano, Via  Celoria 16, I-20133 Milan, Italy\\
andrea.smirne@unimi.it}

\author{Bassano Vacchini}

\address{Università degli Studi di Milano, Dipartimento di Fisica, Via  Celoria 16, I-20133 Milan, Italy\\
Istituto Nazionale di Fisica Nucleare, Sezione di Milano, Via  Celoria 16, I-20133 Milan, Italy\\
bassano.vacchini@mi.infn.it}

\maketitle


\begin{abstract}
In this work, we investigate the non-Markovian features of the Adapted
Caldeira-Leggett model, a computationally efficient framework
recently proposed to capture the essential physics of the standard
Caldeira-Leggett model. While this effective model has been previously
validated for decoherence and einselection, its ability to reproduce
memory effects remains to be explored. By exploiting the model's
capability to explicitly track both system and environment degrees of
freedom, we provide a detailed characterization of non-Markovianity
through the lens of information backflow. We evaluate the buildup of
system-environment correlations and the corresponding modifications of
the environmental state, assessing a quantitative upper bound for
the revival of distinguishability in the reduced dynamics. Our
results, obtained by comparing different distinguishability
quantifiers such as trace distance and the square root of the Jensen-Shannon divergence,
show that while correlations are primarily sensitive to coupling
strength, environmental state changes are more heavily influenced by
temperature. Our analysis substantiates the physical interpretation
of the distinguishability-based approach to non-Markovianity, and
confirms this variant of the Caldeira-Leggett model as a reliable tool for exploring the
microscopic origins of different fundamental phenomena in quantum
mechanics.
\end{abstract}


\section{Introduction}\label{sec:intro}

In most physical scenarios, the primary interest lies in studying the reduced degrees of freedom of a system while accounting for environmental influences through an effective description~\cite{Breuer2002,Rivas2012,Vacchini2024}. Indeed, an exact treatment of all environmental degrees of freedom is often computationally prohibitive, making such effective approaches a necessity.

However, certain situations demand detailed information about the surroundings to ensure a correct physical interpretation. A prominent example is the framework of Quantum Darwinism~\cite{Zurek2009a,Zurek2025,Horodecki2015a,Ryan2021a,Cakmak2021a,Korbicz2021a}, which seeks to explain the emergence of objective classical properties in the quantum evolution through the interaction between a system and its environment. Furthermore, the study of decoherence and its connection to einselection (environment-induced superselection) provides another fundamental case~\cite{Joos2003,Schlosshauer2007,Zurek2025,Zurek2003a,Hornberger2009a,Strasberg2023a}; here, suitable pointer states emerge from the system-environment interaction, leading to classical behavior for specific observables.

Another crucial case arises in the investigation of memory effects in the reduced dynamics~\cite{Rivas2014a,Breuer2012a,Breuer2016a,Devega2017a,Li2018a}, which is the core focus of this paper. Gaining access to the environmental degrees of freedom allows for an exploration of memory effects as arising from information backflow from the bath to the system, as suggested by the Breuer-Laine-Piilo approach to non-Markovianity~\cite{Breuer2009b}.

Given the challenge of tracking all involved degrees of freedom, even numerically, a new effective model has recently been proposed to capture the essential physics of the Caldeira-Leggett~\cite{Caldeira1981a} model while remaining numerically efficient. This Adapted Caldeira-Leggett (ACL) model was introduced by Albrecht~\cite{Albrecht2023a}. While its validity has been demonstrated regarding decoherence and einselection, this work aims to investigate whether the ACL model reliably describes non-Markovian features as well.

Our objective is threefold. First, we test the validity of the ACL model in reproducing a key physical property of the original Caldeira-Leggett framework: the non-Markovianity of the associated reduced dynamics. Second, we exploit the ACL model to investigate the actual information backflow, providing a deeper physical interpretation of the non-Markovian dynamics. Such an analysis has previously been unfeasible due to the aforementioned computational difficulties. Our results provide new insights into the origin of memory effects in the Caldeira-Leggett model and their dependence on the physical parameters characterizing the environment. Finally, we compare the performance of two distinguishability quantifiers, namely trace distance and square root of the Jensen-Shannon divergence, that have proven to have very close performances in the quantification of non-Markovianity, in estimating the information backflow.

The paper is organized as follows. In Sec.~\ref{sec:acl}, we introduce the ACL model, clarifying the details and relevant figures of merit used in environmental modeling. In Sec.~\ref{sec:nm}, we introduce the adopted definition of non-Markovian dynamics, alongside its suggested physical interpretation and the quantities required to validate this perspective. Sec.~\ref{sec:new} details our results for non-Markovianity within the model, benchmarking them against previous work; here, we also discuss the behavior of information backflow and its connection to the establishment of correlations and the system's influence on the environment.
We further benchmark the behavior of trace distance and square root of the Jensen-Shannon divergence in estimating the information backflow.
Finally, in Sec.~\ref{sec:ceo}, we summarize our results and outline potential future developments.

\section{Adapted Caldeira-Leggett model}\label{sec:acl}
The ACL model is a fictitious model of system-environment interaction, designed to describe the dynamics of a continuous variable system interacting with a bosonic environment via position coupling. The model was introduced by Albrecht in Ref.~\citen{Albrecht2023a}, where its effectiveness in describing decoherence and einselection was investigated, and was further studied with respect to its capability to describe equilibration and thermalization in Ref.~\citen{Albrecht2022b}. It arises as an \textit{adaptation} of the well-known Caldeira-Leggett model~\cite{Caldeira1981a,Caldeira1983a,Caldeira1983b} to make it more amenable to numerical analysis. In particular, this model lends itself to an analysis of the time evolution of all involved degrees of freedom, which is an essential ingredient for assessing certain physical features.

In this work, relying on the promising results obtained from the model for the description of the above-mentioned features, we will explore its validity in describing the non-Markovianity of the reduced system dynamics, as well as its interplay with the establishment of correlations between the system and the environment and the feedback of the system on the environment.

\subsection{System and environment modeling}\label{sec:sae}

The ACL model is based on a Hamiltonian of the form
\begin{equation}
\label{eq:Htot}
H = H_S \otimes {1}_E + {1}_S \otimes H_E +\gamma \, q_S \otimes X_E ,
\end{equation}
where the subscripts $S$ and $E$ denote operators acting on the system and environment degrees of freedom, respectively.  
The peculiarity of the model relies on two main ingredients. On the one hand, the system position operator $q_S$ is constructed as a linear combination of creation and annihilation operators of a truncated harmonic oscillator. On the other hand, no specific microscopic structure is assumed for the environmental Hamiltonian $H_E$ and for the environmental operator $X_E$ appearing in the interaction term, which are instead modeled as Hermitian random matrices.  
As a consequence, both the system and the environment are described within finite-dimensional Hilbert spaces.

\begin{figure}[h]
           \includegraphics[width=.49\linewidth]{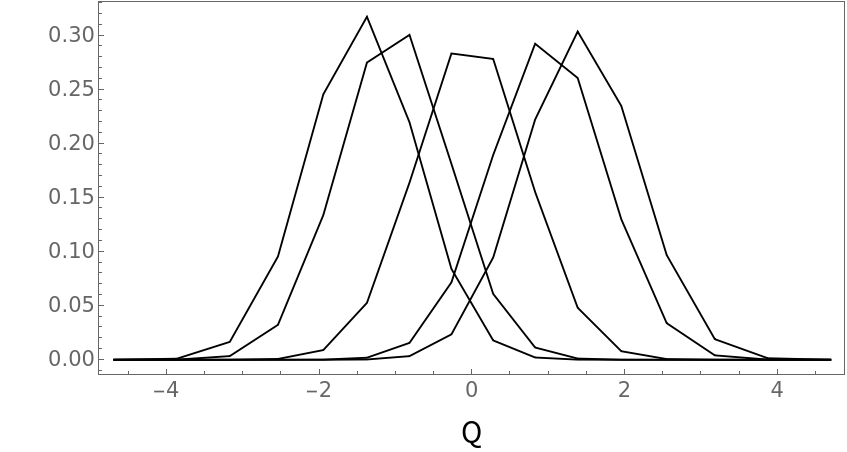}
           \includegraphics[width=.49\linewidth]{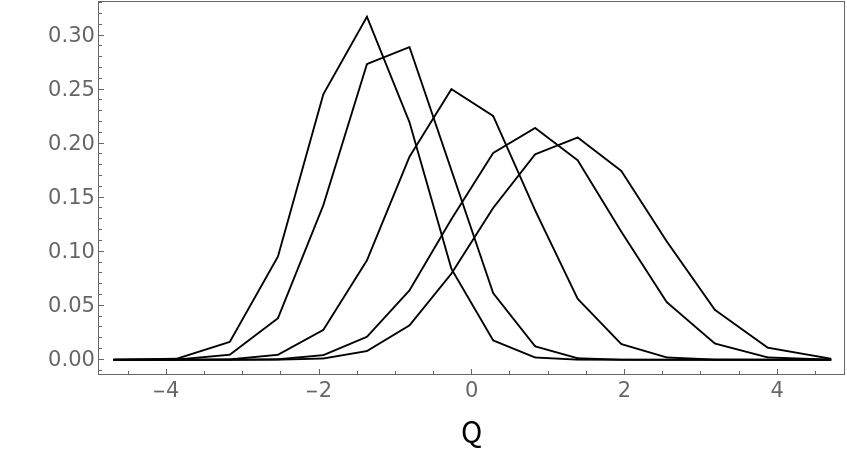}
  \caption{Time evolution of a truncated coherent state  for the free system Hamiltonian $H_S = \omega_S (a_S^\dagger a_S + 1/2)$ (left) and in the presence of damping due to a thermal environment with coupling strength $\cop=0.32$ and bath temperature (see Eq.(\ref{eq:rhoe0})) $\te=0.1$ (right);
  both in units of $\omega_S$. 
  The plots show the squared modulus of the wave-function as a function of position at different times, moving with time from left to right.
  These results benchmark validity of the ACL model for the description of the reduced dynamics.}
  \label{fig:bench}
\end{figure}

The finite dimensionality of the system degrees of freedom is obtained by defining a truncated annihilation operator $a_S$ as
\begin{equation}
\label{eq:troncaa}
   \langle n|a_S|m \rangle = \sqrt{m}\,\delta_{n,m-1},
   \qquad n\in{\{0,\ldots,N_S-1\}}; m\in \{1,\ldots,N_S\}
\end{equation}
where $N_S$ fixes the dimension of the truncated Hilbert space.  
By considering the adjoint operator $a_S^\dagger$, we define their linear combination
\begin{equation}
\label{eq:troncaq}
   q_S= \frac{1}{\sqrt{2}} \left(a_S + a_S^\dagger\right)
\end{equation}
as the dimensionless position operator of the system.  
We note that the definition in Eq.~\eqref{eq:troncaa} implies modified commutation relations,
\begin{equation}
\label{eq:commuta}
\left[a_S, a_S^\dagger \right] = {1}_S - N_S |N_S \rangle \langle N_S| .
\end{equation}
The system Hamiltonian is then defined as $H_S = \omega_S (a_S^\dagger a_S + 1/2)$; in the following we set $\hbar =1$ 
and express all frequencies in units of $\omega_S$.
As discussed in Ref.~\citen{Albrecht2023a}, this definition turns out to be particularly convenient for numerical simulations and reproduces the expected dynamical behavior of truncated coherent states. These states are best obtained~\cite{Opatrny1996a,Albrecht2023a} as truncated coherent superposition of Fock states given by the repeated action of the adjoint of the operator defined in Eq.~\eqref{eq:troncaa}.
The validity of this strategy is illustrated in Fig.~\ref{fig:bench}, where we show the time evolution of a truncated coherent state both for the ideal harmonic-oscillator dynamics and including damping effects due to environmental interaction obtained, demonstrating that the truncation does not introduce spurious effects within the relevant time window.
The environmental Hamiltonian $H_E$ is constructed as a Hermitian random matrix with complex entries.
The dimension of this random matrix is fixed to $N_E$, which sets the effective dimensionality of the environment. The same construction is used for $X_E$.
The choice of the relative sizes of $N_S$ and $N_E$ is optimized so as to ensure numerical feasibility while at the same time allowing the environment to induce the relevant effects on the system dynamics.
The matrices $H_E$ and $X_E$ are independently sampled from the Gaussian unitary ensemble, rescaled by a factor $1/\sqrt{N_E}$ to compensate for the scaling of the eigenvalues with the environmental dimension~\cite{Livan2018}.

In contrast to Ref.~\citen{Albrecht2023a}, the environment in the present work is initially in a mixed state. This necessitated the use of the density matrix formalism from the outset, significantly increasing the computational time. In this regard, our simulations were performed with a system dimension $N_S=16$ and an environment dimension $N_E=64$. These parameters allowed us to ensure that the norm deviation of the truncated coherent states from the exact expression (obtained for the infinite-dimensional case) was limited to $10^{-7}$ for a displacement $\alpha=1$.

\section{Non-Markovian Dynamics and Information Backflow}
\label{sec:nm}

Broadly speaking, the dynamics of an open system is Markovian when memory effects induced by the interaction with the environment 
can be neglected~\cite{Breuer2002}. Starting from this and motivated by the analogous notion in classical stochastic processes~\cite{Feller1971}, 
Markovianity has been extended to the quantum domain in several, generally nonequivalent, definitions~\cite{Rivas2014a,Breuer2016a,Li2018a,Milz2021a}. 
In particular, one can rely on a rigorous and physically transparent way to identify memory effects 
with a bidirectional exchange of information between the open system and the environment. 
In this picture, information flows first from the system to the environment, and subsequently in the opposite direction, so that information about the system’s state at earlier times influences its evolution at later times.
Firstly formulated by using the trace-distance revivals as indicators of information backflow~\cite{Breuer2009b}, 
this approach has then been generalized to include different quantifiers that, as the trace distance, 
evaluate the distinguishability between quantum states, so that a backflow of information manifests 
as an increase in the states' distinguishability~\cite{Vasile2011a,Wissmann2015a,Megier2021a,Smirne2022b}. 

After introducing the general properties needed to identify quantum states' distinguishability quantifiers, we will here briefly recall
how the latter can be used to define and quantify non-Markovianity for open quantum system dynamics.
We will further present a general bound to their revivals, which clarifies the role of system-environment correlations and changes in the environmental state
in establishing information backflow.

\subsection{Distinguishability quantifiers}

Let $\mathfrak{S} (\rho^{(1)}, \rho^{(2)})$ 
be any quantity defined on any couple of quantum states $(\rho^{(1)}, \rho^{(2)})$, i.e., Hermitian, positive operators with trace one,
that satisfies the following properties:
(i.) it is nonnegative, bounded and normalized, according to
  \begin{equation}
    \label{eq:bo} 0 \leqslant \mathfrak{S} (\rho^{(1)}, \rho^{(2)}) \leqslant 1 \quad
    \forall \rho^{(1)}, \rho^{(2)},
  \end{equation}
  with
  \begin{equation}
    \label{eq:norm} \mathfrak{S} (\rho^{(1)}, \rho^{(2)}) = 1 \quad \Leftrightarrow
    \quad {\rho^{(1)} \perp_{\mbox{\footnotesize{supp}}} \rho^{(2)}},
  \end{equation}
  where {$\perp_{\mbox{\footnotesize{supp}}}$} denotes orthogonal supports, and
  \begin{equation}
    \label{eq:indid} \mathfrak{S} (\rho^{(1)}, \rho^{(2)}) = 0 \quad \Leftrightarrow
    \quad \rho^{(1)} = \rho^{(2)};
  \end{equation}
(ii.)  it is contractive under completely positive trace preserving (CPTP) maps:
  \begin{equation}
    \mathfrak{S} (\Lambda [\rho^{(1)}], \Lambda [\rho^{(2)}]) \leqslant \mathfrak{S}
    (\rho^{(1)}, \rho^{(2)}) \quad \forall \rho^{(1)}, \rho^{(2)}, \hspace{0.17em} \hspace{0.17em}
    \forall \text{CPTP } \Lambda; \label{eq:cp}
  \end{equation}
(iii.) it satisfies the triangular inequalities
  \begin{eqnarray}
    \mathfrak{S} (\rho^{(1)}, \rho^{(2)}) -\mathfrak{S} (\rho^{(1)}, \rho^{(3)}) & \leqslant & 
    \mathfrak{S}(\rho^{(2)}, \rho^{(3)}) \quad \forall \rho^{(1)}, \rho^{(2)}, \rho^{(3)} 
    \nonumber\\
    \mathfrak{S} (\rho^{(2)}, \rho^{(1)}) -\mathfrak{S} (\rho^{(3)}, \rho^{(1)}) & \leqslant &
    \mathfrak{S}(\rho^{(2)}, \rho^{(3)}) \quad \forall \rho^{(1)}, \rho^{(2)}, \rho^{(3)}.
    \label{eq:tlike2}
  \end{eqnarray}

These are the requirements we ask for a quantity to be a proper quantifier of the distinguishability between quantum states.
In fact, property (i.) ensures that distinguishability is a non-negative number, it is normalized to 1 (which is useful when comparing different quantifiers), as well as that only and all identical states are indistinguishable 
and that maximal distinguishability is achieved for only and all states that can be discriminated in single-shot measurements.
Property (ii.), which can be seen as an instance of the quantum data-processing inequality~\cite{Nielsen2000}, 
expresses that distinguishability cannot be increased by any operation on a quantum system, 
such as a measurement or an evolution induced by the interaction with a second, initially uncorrelated system in a fixed state.
Finally, property (iii.) is the crucial ingredient to relate the changes in the states' distinguishability of a quantum system
with the information contained within other degrees of freedom;
note that it can be relaxed to a weaker requirement, 
while keeping the same physical picture presented in the next subsections~\cite{Smirne2022b}.

Contractivity under CPTP maps is arguably the most relevant requirement for a quantifier of quantum states' distinguishability,
both because of its physical meaning and because it implies other relevant features. 
In particular, Eq.(\ref{eq:cp}) implies that the quantifier is invariant under
unitary transformations
\begin{equation}
  \label{eq:uinv} \mathfrak{S} (U \rho^{(1)} U^{\dagger}, U \rho^{(2)} U^{\dagger})
  =\mathfrak{S} (\rho^{(1)}, \rho^{(2)}) \quad \forall \rho^{(1)}, \rho^{(2)}, \hspace{0.17em}
  \hspace{0.17em} \forall \text{unitary } U,
\end{equation}
as well as under the tensor product with a fixed state,
\begin{equation}
\mathfrak{S} (\rho^{(1)}, \rho^{(2)}) =\mathfrak{S} (\rho^{(1)} \otimes
  \rho^{(3)}, \rho^{(2)} \otimes \rho^{(3)})\quad \forall \rho^{(1)}, \rho^{(2)}, \rho^{(3)},\label{eq:tens}
\end{equation}
and it further leads to the orthogonality of only and all the pairs of states maximizing the quantifier, 
provided that the invariance of the latter guarantees sufficiency of the map~\cite{Jencova2012a} when restricted to commuting states~\cite{Vacchini2025a}. 

Two significant examples, which we will use in the following analysis, are given by the trace distance $D(\rho^{(1)}, \rho^{(2)})$
and the square root of the Jensen-Shannon divergence $\sqrt{\mathsf{J} (\rho^{(1)}, \rho^{(2)})}$. The former is defined as one half the trace norm of the difference 
between the two quantum states it refers to,
\begin{equation}
  \label{eq:td} D (\rho^{(1)}, \rho^{(2)}) = \frac{1}{2}  \| \rho^{(1)} - \rho^{(2)} \|_1 =
  \frac{1}{2}  \sum_i | \ell_i |,
\end{equation}
where $\| \cdot \|_1$ is the $1 -$norm, so that the $\ell_i$s are the
eigenvalues of the Hermitian traceless operator $\rho^{(1)} - \rho^{(2)}$,
and it then naturally satisfies the properties (i.)-(iii.).
Given the quantum relative entropy~\cite{Bengtsson2017}
\begin{equation}\label{eq:relativa}
  S (\rho, \sigma) = \text{tr} \{ \rho \log \rho \} - \text{tr} \{ \rho \log
  \sigma \},
\end{equation}
the Jensen-Shannon divergence is defined as~\cite{Bengtsson2017,Vacchini2024}
\begin{equation}
 \mathsf{J} (\rho^{(1)}, \rho^{(2)}) = \frac{1}{2 \log 2}  \left( S \left( \rho^{(1)}, \frac{\rho^{(1)} +
  \rho^{(2)}}{2} \right) + S \left( \rho^{(2)}, \frac{\rho^{(1)} + \rho^{(2)}}{2} \right)
  \right),
\end{equation}
and it satisfies property (i.)-(ii.), which are then shared by its square root $\sqrt{\mathsf{J} (\rho^{(1)}, \rho^{(2)})}$;
not only, but the latter is a distance~\cite{Virosztek2021a,Sra2021a}, so that it also satisfies property (iii.). 

\subsection{Non-Markovianity measure}\label{sec:nmma}
Having introduced the notion of quantum-state distinguishability quantifier $\mathfrak{S} (\rho^{(1)}, \rho^{(2)})$, 
we now show how this allows one to define and quantify non-Markovianity for general open quantum system dynamics.

Hence, consider the open system $S$ and the environment $E$, which are uncorrelated at
the initial time $t_0 = 0$, 
\begin{equation}\label{eq:fac}
\rho_{SE} (0) = \rho_S (0) \otimes \rho_E(0), 
\end{equation}
where $\rho_E (0)$ is a fixed environmental state, and which form together a closed system evolving unitarily, 
as fixed by the group of unitary operators $U(t)$. Then, the reduced evolution of the open system
is fixed by a one-parameter family of CPTP maps $\left\{\Lambda (t)\right\}_{t \geq 0}$, according
to~\cite{Breuer2002}
\begin{equation}
  \rho_S (t) = \Lambda (t) [\rho_S (0)] = \text{tr}_E \{ U (t) (\rho_S (0)
  \otimes \rho_E (0)) U^{\dagger} (t) \},
\end{equation}
with $\text{tr}_E$ the partial trace over the environment. 
Given two initial conditions, $\rho^{(1)}_{SE} (0) = \rho^{(1)}_S (0) \otimes \rho_E (0)$ and 
$\rho^{(2)}_{SE} (0) = \rho^{(2)}_S (0) \otimes \rho_E (0)$, so that $\rho^{(i)}_S (t) = \Lambda
(t) [\rho^{(i)}_S (0)]$ for $i=1, 2$, and
instants of time $s$ and $t \geq s$, the difference
\begin{equation}
  \label{eq:delta} \Delta_S \mathfrak{S} (t, s) = \mathfrak{S} (\rho^{(1)}_S
  (t), \rho^{(2)}_S (t)) -\mathfrak{S} (\rho^{(1)}_S (s), \rho^{(2)}_S (s))
\end{equation}
tells us the change of the quantum states' distinguishability
from time $s$ to time $t$, as quantified by $\mathfrak{S}$. 

Now, the basic idea is that a decrease in distinguishability is associated with information flowing out of the open system, 
thus leading to a reduction of the ability to discriminate among the two initial conditions $\rho^{(1)}_S (0)$
and $\rho^{(2)}_S (0)$, while an increase in distinguishability means that some information is flowing into the open system.
The sequence of two time intervals $[t_1, t_2]$ and $[t_2, t_3]$ such that a distinguishability decrease,
$\Delta_S \mathfrak{S} (t_2, t_1)<0$, is followed by an increase, $\Delta_S \mathfrak{S} (t_3, t_2)>0$,
is then interpreted as a bidirectional flow of information:
some of the information about the initial condition that left the open system during $[t_1, t_2]$ is recovered by it during $[t_2, t_3]$.
Such a recovery of information is the essence of the notion of memory effect,
and it can thus be used to define (non-)Markovianity in open system dynamics.
Explicitly, we say that the open-system dynamics $\left\{\Lambda (t)\right\}_{t \geq 0}$ is Markovian when the rate
\begin{equation}
    \sigma(\rho^{(1)}_S(0),\rho^{(2)}_S(0),t) = 
    \frac{d}{d t} \mathfrak{S} (\rho^{(1)}_S(t), \rho^{(2)}_S (t))
\end{equation}
is non-positive for every time and couple of initial states:
\begin{equation}
    \sigma(\rho^{(1)}_S(0),\rho^{(2)}_S(0),t) \leq 0 \quad \forall \,\rho^{(1)}_S(0),\rho^{(2)}_S(0),t.
\end{equation}
Markovian dynamics are those characterized by a unidirectional flow of information out of the open system: any information leaving it 
is irreversibly lost and cannot affect the open system dynamics back again.
Conversely, non-Markovian dynamics are those where there is at least a couple of initial conditions and a time interval
for which some information flows back to the open system, i.e., 
such that $\Delta_S \mathfrak{S} (t, s)>0$ on that interval and then $\sigma(\rho^{(1)}_S(0),\rho^{(2)}_S(0),\tau) >0 $
for some $\tau \in [s, t]$.

Not only does this approach allow for a clear-cut definition of quantum Markovianity, 
but it also provides a natural way to quantify non-Markovianity. 
The first step to do so is to integrate
over all the increases of distinguishability, i.e. over all the information about the initial conditions 
that flows back to the open system
in the course of the evolution, thus getting
\begin{equation}\label{eq:NM0}
    \mathcal{N}(\rho^{(1)}_S(0),\rho^{(2)}_S(0)) =  \int_{\sigma >0} dt \, \sigma(\rho^{(1)}_S(0),\rho^{(2)}_S(0),t).
\end{equation}
By further maximizing over all couples of initial conditions,
\begin{equation}\label{eq:NMm}
    \mathcal{N}_{\Lambda(t)} = \max_{(\rho^{(1)}_S(0),\rho^{(2)}_S(0))} \int_{\sigma >0} dt \, \sigma(\rho^{(1)}_S(0),\rho^{(2)}_S(0),t),
\end{equation}
we are left with a quantity that is associated with the family of CPTP maps fixing the dynamics of the open system (as stressed by the subscript $\Lambda(t)$),
without depending on a specific choice of the initial conditions.
Note that, at least for some distinguishability quantifiers including the trace distance, the evaluation of the maximum in Eq.(\ref{eq:NMm}) can be strongly simplified by general results~\cite{Wissmann2012a,Liu2014b,Wissmann2015a}, implying in particular that the optimal pair is always given by two orthogonal states on the boundary of the state space.

\subsection{Bounds on system-environment information exchange}

The physical picture based on the notion of information flow can be further substantiated by taking into account the dynamical evolution of the information, not only inside the open system, but also outside it and then at the global level.

Let thus (where for the sake of readability we leave the dependence on $(\rho^{(1)}_S(0), \rho^{(2)}_S (0))$ implied)
\begin{equation}
\mathcal{I}_{int}(t) = \mathfrak{S} (\rho^{(1)}_S(t), \rho^{(2)}_S (t))
\end{equation}
be the amount of information within the open system, i.e., that can be accessed via measurements on $S$ only,
and
\begin{equation}
\mathcal{I}_{ext}(t) = \mathfrak{S} (\rho^{(1)}_{SE}(t), \rho^{(2)}_{SE} (t)) - \mathfrak{S} (\rho^{(1)}_S(t), \rho^{(2)}_S (t))
\end{equation}
the amount of information outside it.
The factorized initial condition and Eq.(\ref{eq:tens}) imply
$$
\mathfrak{S} (\rho^{(1)}_{SE}(0), \rho^{(2)}_{SE} (0)) = \mathfrak{S} (\rho^{(1)}_{S}(0), \rho^{(2)}_{S} (0)),
$$
so that $\mathcal{I}_{ext}(0)=0$: since the global initial condition is a factorized state, with a fixed state of the environment, 
all the initial information about it is contained within the open system. 
Moreover, the global unitary evolution does not modify the total amount of information, i.e., Eq.(\ref{eq:uinv})
implies
$$
\mathfrak{S} (\rho^{(1)}_{SE}(t), \rho^{(2)}_{SE} (t)) = \mathfrak{S} (\rho^{(1)}_{SE}(0), \rho^{(2)}_{SE} (0)),
$$
so that all in all we have
\begin{equation}
    I_{int}(t) + I_{ext} (t) = I_{int}(0). 
\end{equation}
The information about the initial condition, initially only within the open system, is then exchanged between the open system and the environment, so that any backflow of information to the open system in the time interval $[s,t]$ corresponds to a decrease of the external information that has been collected up to time $s$. 
By using properties (ii.) and (iii.), 
one can upper bound the distinguishability increase due to a backflow of information, according to
\begin{align}
    \Delta_S \mathfrak{S} (t, s) \leqslant I_{ext} (s) \leqslant 
    & \mathfrak{S}(\rho^{(1)}_{SE} (s),
  \rho^{(1)}_S (s) \otimes \rho^{(1)}_E (s)) + 
  \mathfrak{S}(\rho^{(2)}_{SE} (s), \rho^{(2)}_S (s) \otimes \rho^{(2)}_E (s)) \nonumber \\
  &  + \mathfrak{S}(\rho^{(1)}_{E} (s), \rho^{(2)}_{E} (s)).\label{eq:main}
\end{align}
This bound provides us with a general characterization of the
microscopic mechanisms behind the occurrence of memory effects, extending the corresponding physical picture 
for the trace distance~\cite{Laine2010a,Smirne2011a,Mazzola2012a,Smirne2013b,Laine2010b,Breuer2016a,Campbell2019b,Amato2018a} to general distinguishability quantifiers~\cite{Smirne2022b}.
The amount of states' distinguishability that can be recovered in a certain time interval $[s,t]$ is upper bounded by the information that is outside the open system at time $s$. Such external informal is in turn bounded by the sum of the total correlations, both classical and quantum, that are present in the two global states that evolve from the initial conditions to time $s$, and the distinguishability in the corresponding environmental states. Indeed, the external information is contained not only within the environment, but also in the system-environment correlations. Conversely, Eq.(\ref{eq:main}) implies that any backflow of information is induced by the correlations established between the system and the environment, and by the information stored in the environment.

Of course, different distinguishability quantifiers $\mathfrak{S}(\rho^{(1)}_{S}, \rho^{(2)}_{S})$ will generally lead to different degrees of non-Markovianity, possibly even to a different classification of Markovian and non-Markovian dynamics, as well as to a different quantification of the system-environment correlations and changes in the environmental states; the comparison among two relevant quantifiers is indeed a central subject of our investigation. On the other hand, we remark that the physical picture now illustrated does hold irrespectively of the considered quantifier.

\section{Results}
\label{sec:new}

We are now ready to present the results of our investigation, namely, the assessment of the degree of non-Markovianity and its connection with the system-environment correlations and changes in the environmental states for the ACL.

\subsection{Time-evolution of the distinguishability quantifiers}\label{sec:teo}

\begin{figure}[h]
        \includegraphics[width=.49\linewidth]{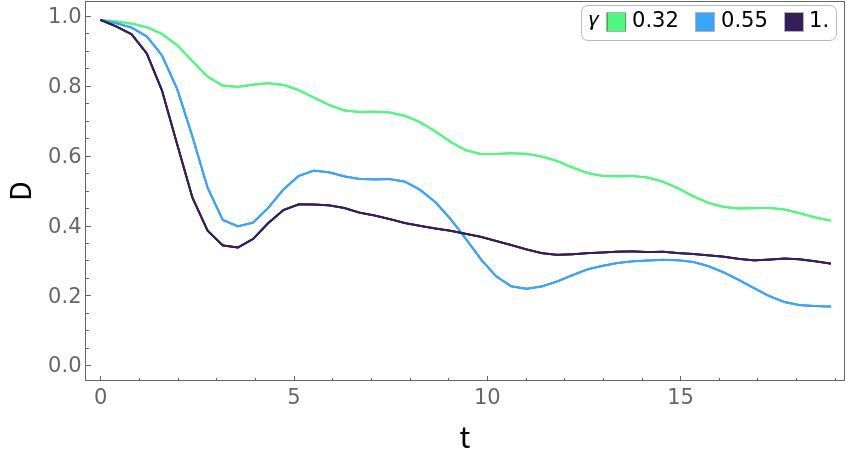}
         \includegraphics[width=.49\linewidth]{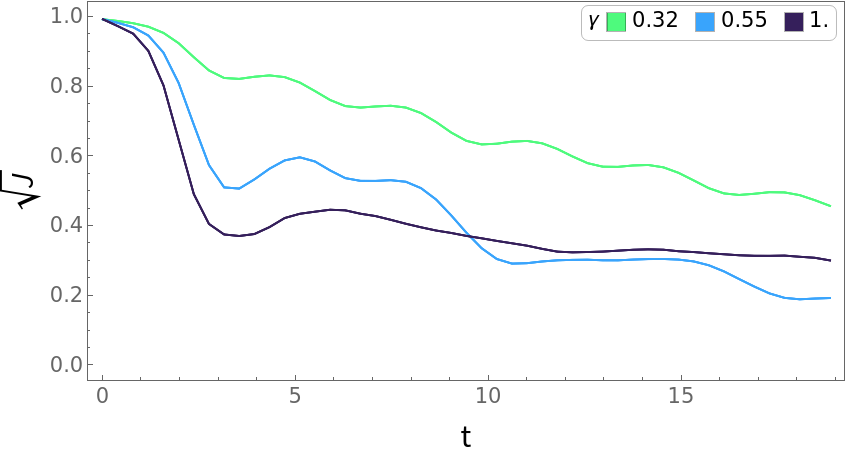}
 \caption{Trace distance $D(\rho^{(1)}_S(t),\rho^{(1)}_S(t))$ (left) 
 and square root of the Jensen-Shannon divergence $\sqrt{\mathsf{J} (\rho^{(1)}(t), \rho^{(2)}(t))}$ (right)
 as a function of time, for different values of the coupling $\cop$, at fixed bath temperature $\te = 1$; 
 the two initial open-system states are as in Eq.(\ref{eq:initcc}) for $\alpha=1$.%
}\label{fig:td}
\end{figure}

As explained in Sec.\ref{sec:nmma}, the notion of non-Markovianity investigated in this work is based on the time evolution 
of state distinguishability, obtained by comparing the dynamics arising from two distinct initial open-system states, 
assuming initial global product states as in Eq.(\ref{eq:fac}).
Throughout our analysis, we fix the initial system states as the truncated coherent states
\begin{equation}\label{eq:initcc}
    \rho_S^{(1)}(0) = \ket{\alpha}\bra{\alpha}, \quad \rho_S^{(2)}(0) = \ket{-\alpha}\bra{-\alpha},
\end{equation}
defined as detailed in Sec.~\ref{sec:sae}, considering $\alpha=1$, while the environment is taken to be initially in a Gibbs state at temperature $\te$ (we set $k_B = 1$)
\begin{equation}\label{eq:rhoe0}
    \rho_E(0) = \frac{e^{-H_E/\te}}{Z}, \qquad Z = \text{tr}\left\{e^{-H_E/\te}\right\}.
\end{equation}

In Fig.~\ref{fig:td}, we show the time evolution of two distinguishability quantifiers: 
the trace distance, $\mathfrak{S} (\rho_S^{(1)}(t), \rho_S^{(2)}(t)) \mapsto D(\rho^{(1)}_S(t),\rho^{(2)}_S(t))$ (left panel),
and the square root of the Jensen-Shannon divergence, 
$\mathfrak{S} (\rho_S^{(1)}(t), \rho_S^{(2)}(t)) \mapsto \sqrt{\mathsf{J} (\rho_S^{(1)}(t), \rho_S^{(2)}(t))}$ (right panel),
as functions of time, for different values of the system-environment coupling strength $\cop$ and fixed environmental temperature $\te$.
For both quantifiers and across the entire parameter regime considered,
we observe clear non-monotonic time evolutions, 
indicating the widespread presence of memory effects in the model. 
For a small coupling, 
the dynamics exhibits repeated oscillations with relatively small amplitudes, corresponding to modest but persistent revivals of distinguishability 
over the full time window taken into account.
As the coupling strength increases, more pronounced revivals appear, but which now tend to concentrate into two distinct time intervals. 
Moreover, we note that the first dip of quantum states' distinguishability and subsequent revival is more pronounced as quantified by the trace distance
compared to the square root of the Jensen-Shannon divergence.
For the strongest coupling, the first revival is further amplified, while the second one is almost completely suppressed. 
This behavior reflects a competition between short-time information backflow, which is strengthened by increasing the coupling, 
and an accelerated relaxation towards a (quasi-)\footnote{Since the global system considered is finite-dimensional, the open-system state does not converge asymptotically to a stationary state, but recurrences occur on sufficiently long time scales~\cite{Bocchieri1957a}.} steady state, 
in which oscillations are largely diminished.

\subsection{Degree of non-Markovianity}

We now move to the quantifier of the degree of non-Markovianity $\mathcal{N}$ defined in Eq.(\ref{eq:NM0}) (implying the dependence on the two initial conditions in Eq.(\ref{eq:initcc})).
By integrating the revivals of distinguishability, $\mathcal{N}$ provides a cumulative assessment of the information backflow, 
and thus of the memory effects over the entire evolution. This allows us to compare different dynamical scenarios, 
such as the presence of many small revivals distributed over time versus a smaller number of more pronounced revivals. 
We focus on a fixed pair of initial states and therefore evaluate $\mathcal{N}$, rather than the fully optimized non-Markovianity measure $\mathcal{N}_{\Lambda(t)}$
defined in Eq.(\ref{eq:NMm}), since the latter would require an optimization over initial states that is computationally prohibitive given the high dimensionality 
of the open system; in any case, $\mathcal{N}$ provides a lower bound to $\mathcal{N}_{\Lambda(t)}$.
In our numerical analysis, the integral in Eq.~(\ref{eq:NM0}) is evaluated up to a finite maximum time~$T_{\max}$, for all parameters considered. 
Furthermore, the integral is computed using a discretized time grid; we verified convergence by varying the time step $\Delta t$ and confirming that doubling it leads to no appreciable change in the resulting value of the quantifier.
We report results only for the trace distance, since those obtained using the square root of the Jensen-Shannon divergence are qualitatively and largely quantitatively similar.

\begin{figure}[h]
         \includegraphics[width=.49\linewidth]{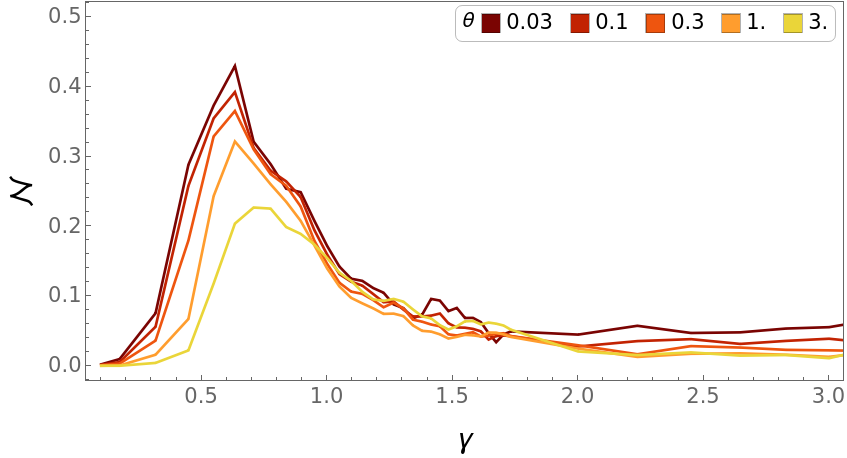}
         \includegraphics[width=.49\linewidth]{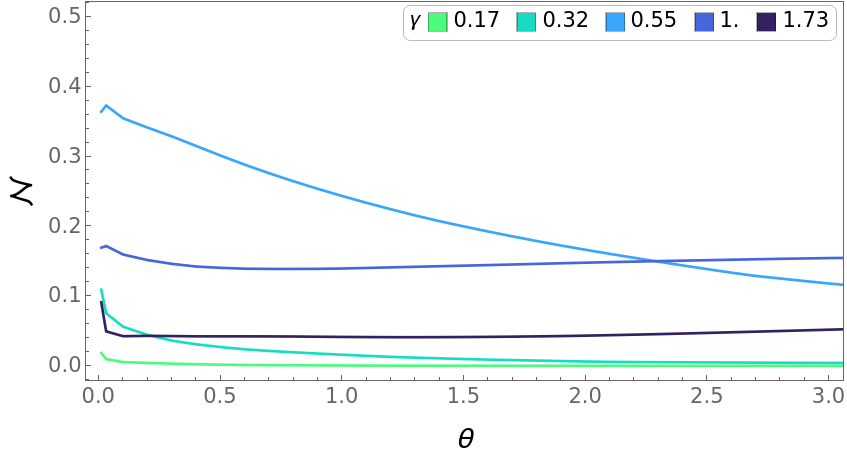}
  \caption{Non-Markovianity $\mathcal{N}$ as defined in Eq.(\ref{eq:NM0}) for the two initial states in Eq.(\ref{eq:initcc})
  with the trace distance as distinguishability quantifier, as a function of the coupling $\cop$ for different values of $\te$ (left) 
  and as a function of the temperature $\te$ for different values of $\cop$ (right).}\label{fig:nm}
\end{figure}

In Fig.~\ref{fig:nm}, we show the behavior of $\mathcal{N}$ as a function of the coupling strength $\cop$ for different values of the temperature $\te$ (left panel), and as a function of~$\te$ for different values of $\cop$ (right panel).
As a function of the coupling, non-Markovianity displays a non-monotonic behavior, reflecting the competition between the physical mechanisms discussed above. Increasing the coupling from the small- to the intermediate-coupling regime enhances the exchange of information between the open system and the environment over the entire evolution, 
while further moving to stronger couplings the accelerated relaxation toward equilibrium prevails:
distinguishability oscillations are progressively suppressed, and eventually the entire backflow of information, as quantified by $\mathcal{N}$, is significantly reduced.
Increasing the temperature $\te$ lowers the maximum value of $\mathcal{N}$ as a function of $\cop$, 
while shifting its position towards larger values of $\cop$,
and for strong couplings $\te$ has a limited impact on $\mathcal{N}$. 

We stress that both the non-monotonic dependence of $\mathcal{N}$ on $\cop$, with a maximum in the intermediate-coupling regime, 
and the tendency towards a temperature-independent behavior at strong coupling are also observed 
in the non-Markovianity measure of the standard Caldeira-Leggett model studied in Ref.~\citen{Einsiedler2020a}, 
where the Bures distance was employed to quantify distinguishability.
Beyond confirming the robustness of the information-backflow approach to non-Markovianity with respect to the specific choice of distinguishability quantifier, this agreement supports the ability of the ACL model to capture the essential physical features of quantum Brownian motion, also for what non-Markovianity is concerned. 

This conclusion is further corroborated by looking at the dependence of $\mathcal{N}$ on the temperature $\te$.
As for the Caldeira-Leggett model~\cite{Einsiedler2020a}, the low-temperature regime is that showing the highest degree of non-Markovianity,
even tough, for certain values of the coupling strength, zero temperature does not correspond to the absolute maximum of $\mathcal{N}$, 
but a weak non-monotonic dependence on~$\te$ can be observed. 
Differently from the Caldeira-Leggett model, instead, we see that 
strongly increasing the temperature can actually enhance non-Markovianity, for large values of the coupling.
We attribute this behavior to the finite size of the environment considered here: higher bath temperatures lead to stronger impact of the truncation procedure described in Sec.~\ref{sec:sae} due to the larger occupation of the highest levels taken into account, which turns into finite-size effects amplifying memory effects.

\subsection{Role of system-environment correlations and environmental evolution in information backflow}

   
While the analysis of quantifiers such as $\mathcal{N}$ provides a characterization of the non-Markovianity of the reduced open-system dynamics,
identifying the microscopic origin of such non-Markovianity requires examining the time evolution of global quantities, 
accounting for the dynamics within the environment
and the buildup of system-environment correlations induced by the interaction.
The bound in Eq.(\ref{eq:main}) enables precisely this investigation, by linking revivals in the distinguishability of open-system states 
to the establishment of total correlations
and to changes in the environmental state, all expressed in terms of the same distinguishability quantifier $\mathfrak{S}(\rho^{(1)},\rho^{(2)})$.

\begin{figure}[h!]
    \includegraphics[width=.49\linewidth]{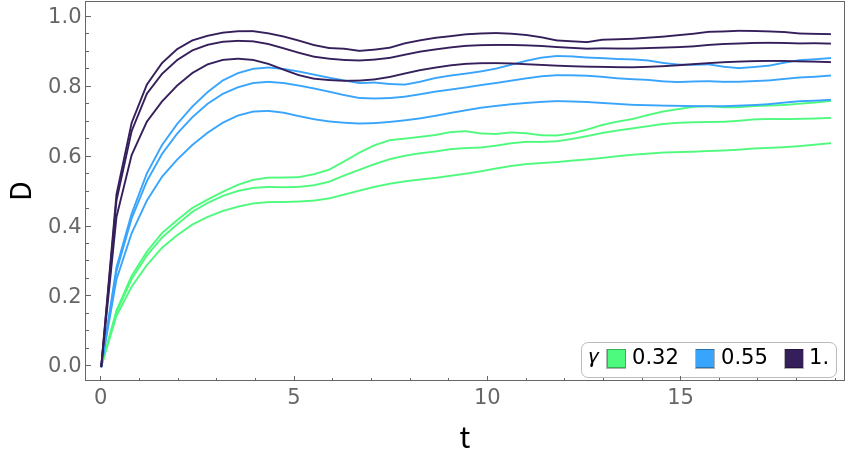}
         \includegraphics[width=.49\linewidth]{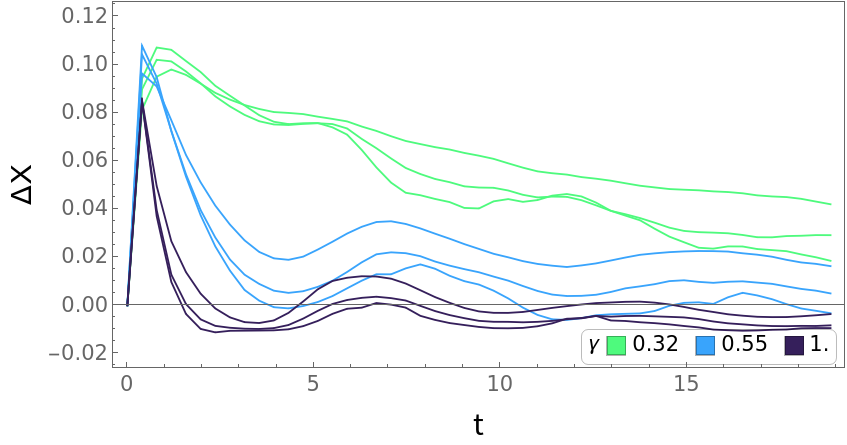}
         \includegraphics[width=.49\linewidth]{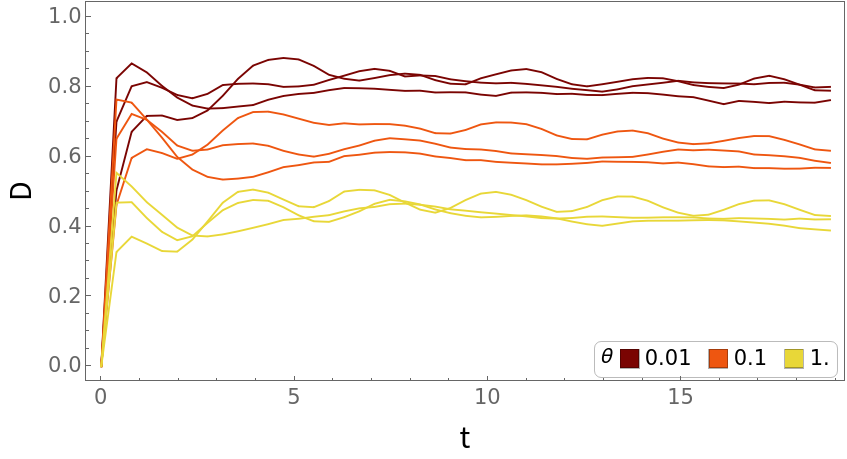}
         \includegraphics[width=.49\linewidth]{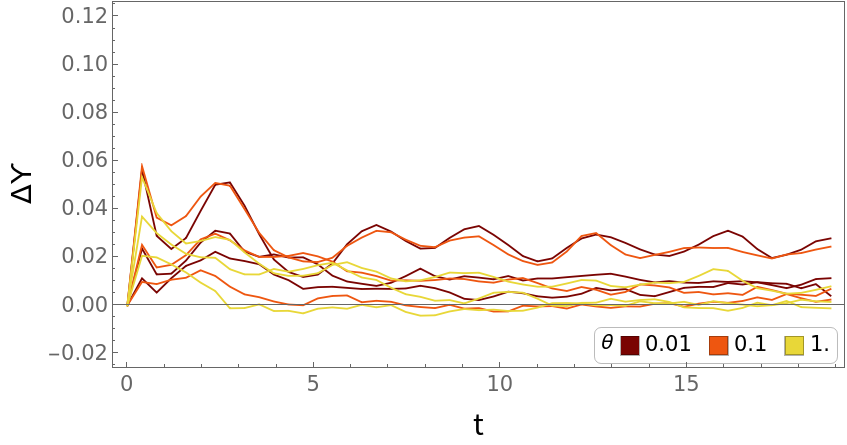}
  \caption{(Left panels) Trace-distance system-environment correlations $D(\rho^{(1)}_{SE}(t),\rho^{(1)}_S(t)\otimes\rho^{(1)}_E(t))$ (upper panel)
  and changes in the environmental state $D(\rho^{(1)}_{E}(t),\rho^{(2)}_E(t))$ (lower panel) as a function of time,
for different values of  coupling and temperature $\cop = \{0.32, 0.55, 1\}$; $\te =\{0.01, 0.1, 1\}$; the other parameters are as in Fig.~\ref{fig:nm}.
(Right panels) Difference between the system-environment correlations evaluated with the square root of the Jensen-Shannon divergence and those evaluated via the trace distance, $\Delta X = \sqrt{\mathsf{J} (\rho^{(1)}_{SE}(t),\rho^{(1)}_S(t)\otimes\rho^{(1)}_E(t))}- D(\rho^{(1)}_{SE}(t),\rho^{(1)}_S(t)\otimes\rho^{(1)}_E(t))$, (upper panel) and difference between the changes in the environmental states with the two quantifiers, 
$\Delta \Upsilon = \sqrt{\mathsf{J} (\rho^{(1)}_{E}(t),\rho^{(2)}_E(t))}- D(\rho^{(1)}_{E}(t),\rho^{(2)}_E(t))$, (lower panel) for the same parameters as the left panels. 
In the upper panels, the different colors refer to distinct values of the coupling, while in the lower panels they refer to distinct values of the temperature;
this reflects the larger impact of coupling and temperature variations on the evolution of, respectively, system-environment correlations and changes in the environmental state. In the upper panels, for a fixed coupling, the temperature increases moving from upper to lower lines (referring to the values at the first maximum),
while in the lower, for a fixed temperature, the coupling increases moving from lower to upper lines. Note the difference in scale between the right and the left panels.}\label{fig:bound1}
\end{figure}

In the left panels of Fig.~\ref{fig:bound1}, we report the time evolution of the total correlations, $D(\rho^{(1)}_{SE}(t),\rho^{(1)}_S(t)\otimes\rho^{(1)}_E(t))$,
and of the distinguishability in the environmental states $D(\rho^{(1)}_{E}(t),\rho^{(2)}_E(t))$, both quantified using the trace distance,
for different values of the coupling strength $\cop$ and the temperature $\te$;
for the choice of initial conditions in Eq.(\ref{eq:initcc}), the dynamics of the total correlations of $\rho^{(1)}_{SE}(t)$ and $\rho^{(2)}_{SE}(t)$ coincide.
The interaction progressively establishes system-environment correlations, 
which build up more rapidly and reach larger values for stronger coupling and lower temperature; indeed, variations in the coupling strength have a more pronounced effect on the evolution of global correlations than changes in temperature.
We highlight these features in the left panels of Fig.~\ref{fig:bound1} by using colors to emphasize the bunching of lines around the values of the physical parameters that most significantly affect each figure of merit: namely, the coupling strength for the correlations and the temperature for the environmental changes.
Across all regimes considered, the total correlations tend to saturate at long times, as the system approaches a (quasi-)stationary state. 
Apart from modifying the speed at which the system-environment correlations are formed, both~$\cop$ and~$\te$ 
have a relatively limited impact on their oscillations. 

Both the distinguishability of environmental states and the system-environment correlations quickly approach a first peak, which grows with $\cop$ and shrinks with $\te$,
where now the temperature has a more significant impact than the coupling strength.
Interestingly, here we can observe a behavior that mirrors the competing mechanisms in the backflow of information discussed in the previous section: 
while increasing the coupling enhances the magnitude of changes in the environmental state, it also suppresses their oscillations at intermediate and long time scales. 

Overall, the information outside the open-system leading to the backflow of information at the origin of non-Markovianity
is shared by the system-environment correlations and the environmental state. The information within the latter is predominant at very short times, 
rapidly reaching a local maximum and then oscillating around it, whereas the information within correlations continues to grow throughout the dynamics, 
becoming prevalent at longer times. However, while correlations almost saturate to the maximal possible value, environmental changes have a smaller impact in absolute terms.

These behaviors are shared qualitatively and mostly quantitatively by the corresponding quantities,
$\sqrt{\mathsf{J} (\rho^{(1)}_{SE}(t),\rho^{(1)}_S(t)\otimes\rho^{(1)}_E(t))}$ and $\sqrt{\mathsf{J} (\rho^{(1)}_{E}(t),\rho^{(2)}_E(t))}$, as evaluated via the square root of the Jensen-Shannon divergence. For this reason, in the right panel of Fig.~\ref{fig:bound1} 
we report the difference between the Jensen-Shannon-based and trace-distance-based quantifiers of, respectively, system-environment correlations and environmental distinguishability. This difference is most pronounced at short times, where in particular for system-environment correlations it exhibits an initial peak that is almost independent of temperature and coupling strength, before decaying towards smaller values, with a faster decay for increasing coupling. A similar trend is observed for the difference between the environmental distinguishability quantifiers, although in this case the initial peak is less pronounced. We further note that, although the square root of the Jensen-Shannon divergence typically yields larger estimates than the trace distance, the differences for both system-environment correlations and environmental distinguishability can take on negative values, which demonstrates that no strict hierarchy exists between the two distinguishability quantifiers.

\begin{figure}[h!]
        \includegraphics[width=.49\linewidth]{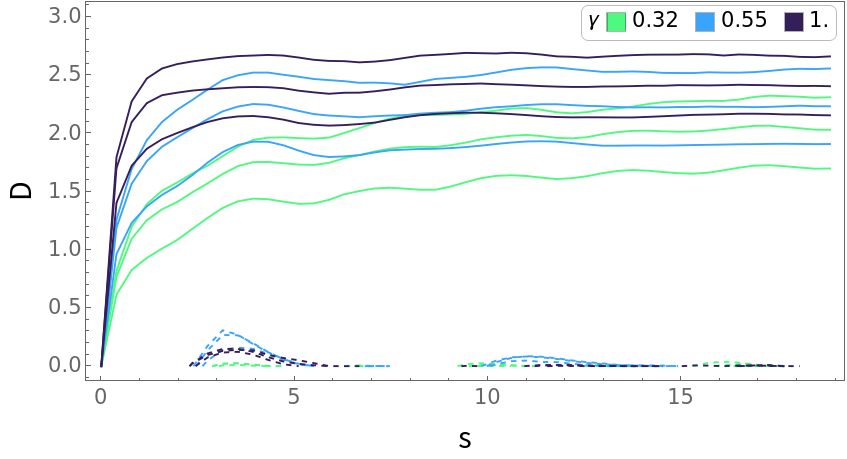}
         \includegraphics[width=.49\linewidth]{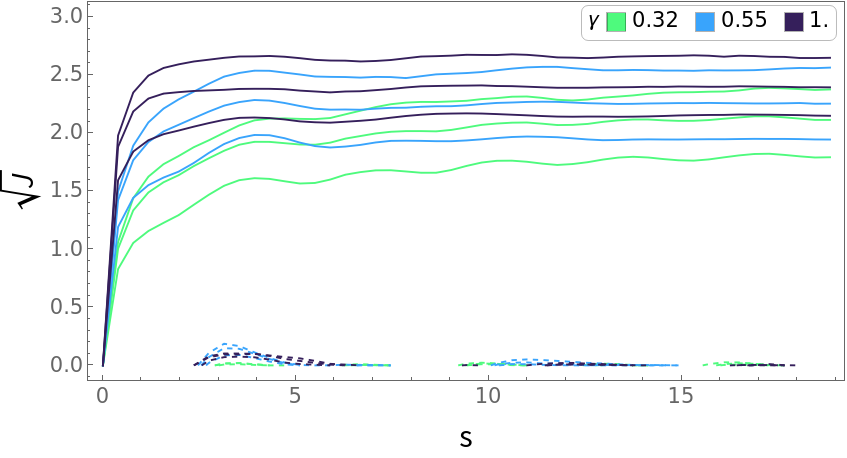} 
  \caption{System-environment correlations plus changes in the environmental state, i.e., r.h.s. of the bound in Eq.(\ref{eq:main}), (solid lines)
  and variation of the open-system states' distinguishability, i.e., l.h.s. of Eq.(\ref{eq:main}), (dashed lines)
  as a function of the intermediate time $s$,
  for the trace distance, $\mathfrak{S} (\rho^{(1)}, \rho^{(2)}) \mapsto  D(\rho^{(1)},\rho^{(2)})$ (left panel),
and the square root of the Jensen-Shannon divergence, 
$\mathfrak{S} (\rho^{(1)}, \rho^{(2)}) \mapsto \sqrt{\mathsf{J} (\rho^{(1)}, \rho^{(2)})}$ (right panel); 
for each $s$, the final time $t$ for the variation of distinguishability is taken as the closest local
maximum (see Fig.~\ref{fig:td}).}\label{fig:bound2}
\end{figure}

Finally, the combined behavior of system-environment correlations and changes in the environmental state, that is the r.h.s. of the bound in Eq.(\ref{eq:main}),
is shown in Fig.~\ref{fig:bound2},
along with the corresponding variation of the open-system states' distinguishability, that is the l.h.s. of Eq.(\ref{eq:main}), evaluated using both the trace distance and the square root of the Jensen-Shannon divergence quantifiers.
In both cases, even though the applicability of the bound given by the r.h.s. of Eq.(\ref{eq:main}) for estimation of the
l.h.s. is indeed limited, the bound is nevertheless able to qualitatively reproduce some features of the dynamics. In particular, the non-monotonic dependence on the coupling strength is captured, at least for intermediate and high temperatures, as is the monotonic decrease with increasing temperature.
We also note that the behavior of the r.h.s. of Eq.(\ref{eq:main}) for the two distinguishability quantifiers is more similar than the behavior of the l.h.s.:
while the open-system states trace distance exhibits a more pronounced first revival than the square root of the Jensen-Shannon divergence 
(see the discussion at the end of Sec.~\ref{sec:teo}), such a feature is essentially absent in the corresponding upper bounds.
Again, we observe that the upper bounds for the two distinguishability quantifiers are qualitatively and mostly quantitatively similar,
with the more significant differences localized in the short-time regime.
Overall, these results are in agreement with previous findings in exactly solvable models, in which both quantifiers demonstrated very similar performance~\cite{Megier2021a,Megier2022a,Vacchini2024a}. Notably, our analysis confirms a slightly higher sensitivity of the trace distance in detecting revivals of distinguishability.

\section{Conclusions}
\label{sec:ceo}

In this manuscript, we have explored the validity of the ACL model in characterizing the non-Markovianity of the Caldeira-Leggett model, focusing specifically on the information backflow that is considered to be at the heart of non-Markovian behavior. By exploiting the ability of the ACL model to conveniently address both system and environment degrees of freedom, we have evaluated the amount of correlations established between the system and its surroundings, alongside the modifications in the environmental state induced by the interaction.

These quantities allow us to investigate a quantitative upper bound for the revival of distinguishability between system states, which is used as a criterion to identify non-Markovian dynamics. Our results indicate that, while correlations are sensitive to the coupling strength and weakly dependent on the environmental temperature, the opposite holds true for changes in the environmental states, which are more heavily influenced by temperature. These estimates appear to be consistent regardless of the chosen distinguishability measure, as we ascertained by comparing the trace distance and the square root of the Jensen-Shannon divergence.

Our results confirm that the ACL model is a reliable and
computationally efficient proxy for the Caldeira-Leggett dynamics,
particularly for exploring the emergence of memory effects. By
establishing a quantitative connection between system-environment
correlations and information backflow, we provide a solid basis for
the physical interpretation of non-Markovianity. Future work will test
the generality of these findings across diverse environmental
geometries, including discrete spin-environments, to further
consolidate our understanding of quantum memory.

\section*{Acknowledgments}

This work has been supported by the Italian Ministry of Research and Next Generation EU via the
PRIN 2022 project Quantum Reservoir Computing
(QuReCo) (contract n. 2022FEXLYB), and
the NQSTI-Spoke1-BaC project QuSynKrono (contract n.
PE00000023-QuSynKrono). We gratefully acknowledge the computing resources provided by AMiCO (\url{http://amico.mi.infn.it}), an opportunistic resource cluster operated by the IT service of the Physics Department of Università degli Studi and INFN Milano - Italy.

\section*{ORCID}

\noindent Luciano Manara - \url{http://orcid.org/0000-0001-9477-9497}

\noindent Andrea Smirne - \url{https://orcid.org/0000-0003-4698-9304}

\noindent Bassano Vacchini - \url{https://orcid.org/0000-0002-7574-9951}

\renewcommand\bibname{References}

\end{document}